\documentclass[11pt,notoc]{JHEP3}

\usepackage{epsfig}
\usepackage{mathrsfs}
\usepackage{amsmath,amssymb}
\usepackage{graphicx}
\usepackage{amsmath}

\newcommand{\bea}{\begin{eqnarray}}
\newcommand{\eea}{\end{eqnarray}}
\newcommand{\be}{\begin{equation}}
\newcommand{\ee}{\end{equation}}
\newcommand{\p}{\partial}

\newcommand{\ov}{\over}
\newcommand{\dd}{\textrm{d}}

\def\ben{\begin{equation}}
\def\een{\end{equation}}

    \let\p=\partial

\let\w=\omega

\let\pa=\partial
\def\be{\begin{equation}}
\def\ee{\end{equation}}
\def\ba{\begin{array}}
\def\ea{\end{array}}

\def\dd{\textrm{d}}

\title{A Holographic model for Non-Relativistic Superconductor}

\author{ Shi Pu\footnote{shipu@th.physik.uni-frankfurt.de}\\
Department of Modern Physics,\\
 University of Science and Technology of China, Anhui 230026, China\\

Institut f\"ur Theoretische Physik,\\
 Johann Wolfgang Goethe-Universit\"at, Max-von-Laue-Str. 1, D-60438,\\
 Frankfurt am Main, Germany}

\author{ Sang-Jin Sin\footnote{sangjin.sin@gmail.com}\\
Department of Physics, Hanyang University,\\
 Seoul 133-791, Korea\\}

\author{ Yang Zhou\footnote{yzhou@itp.ac.cn}\\
Institute of Theoretical Physics\\
Chinese Academic of Science\\
Beijing 100190, PRC\\

Kavli Institute for Theoretical Physics China at the Chinese Academy of Sciences(KITPC-CAS)\\

Key Laboratory of Frontiers in Theoretical Physics£¬Institute of Theoretical Physics£¬Chinese Academy of Sciences}

\abstract{We build a holographic description of non-relativistic  system for superconductivity in strongly interacting
condensed matter via gauge/gravity duality. We focus on the phase transition and give an example to show that a simple
gravitational theory can provide a non-relativistic holographical dual description of a superconductor. There is also a
critical temperature like the relativistic case, below which a charged condensation field appears by a second order
phase transition and the (DC) conductivity becomes infinite. We also calculated the frequency dependent conductivity. }


\begin{document}

\tableofcontents

\section{Introduction and Summary}
AdS/CFT correspondence is one of the most important results from string theory \cite{Maldacena:1997re}, which maps a
conformal gauge field theory on the boundary to string theory in asymptotically anti de Sitter spacetimes.
%
A semi-classic version of this duality has appeared as gauge/gravity duality  and such duality has become a powerful tool to understand the strongly coupled gauge theory and it
was extended to describe aspects of strongly coupled QCD such as properties of quark gluon plasma in heavy ion collisions
at RHIC \cite{son,SZ,Mateos:2007ay} and hadron physics.

More recently, it has been attempted to use this correspondence  to
describe  certain condensed matter systems  such as the Quantum Hall
effect \cite{Hartnoll:2007ai} , Nernst effect \cite{Hartnoll:2007ih,
Hartnoll:2007ip, Hartnoll:2008hs}, superconductor
~\cite{Hartnoll:2008vx}~\cite{Ammon:2009fe}~\cite{Horowitz:2008bn}
and FQHE (fractional quantum hall effect) ~\cite{Fujita:2009kw}. All
of these phenomena have dual gravitational descriptions. As pointed
in ~\cite{Herzog:2008wg}, there is a large class of interesting
strongly correlated electron and atomic systems that can be created
and studied in experiments. In some special conditions, these
systems exhibit relativistic dispersion relations, so the dynamics
near a critical point is well described by a relativistic conformal
field theory. It is expected that such field theories which can be
studied holographically have dual AdS geometries. To describe more
non-relativistic condensed matter systems, this duality has even
been extended to non-relativistic conformal field theory which has
Schr$\ddot{\textrm{o}}$dinger symmetry ~\cite{Herzog:2008wg}.

In condensed matter systems,   traditional theories are based on two themes. One is Landau Fermi liquid theory and
the other is symmetry breaking. High $T_c$ superconductor is a  phenomena waiting for a new theory. Conventional superconductors are well described by BCS theory \cite{parks} while  some basic aspects
of unconventional superconductors, including the pairing mechanism, remain  to be understood. There are many hints that the normal state in these materials can not be described by the
standard Fermi liquid theory \cite{hightc} and many of  unconventional superconductors, such
as the cuprates and organics, are layered and much of the physics is 2+1 dimensional.

In ~\cite{Hartnoll:2008vx}, a model of a strongly coupled system which develops superconductivity was developed based
on the holography, which is an Abelian-Higgs model in a warped space time. While the electrons in real materials are
non-relativistic, the model in ~\cite{Hartnoll:2008vx} is for relativistic system. Therefore it is natural to ask
whether one can develop a similar theory with non-relativistic kinematics. The purpose of this paper is to answer this
question. The boundary field theory in our model is 2+1 dimensional. However, due to the structure of non-relativistic
AdS/CFT correspondence \cite{Son:2008ye}, bulk  theory of our model should be 4+1 dimensional.
We use a complex scalar field to describe the charged
condensation field. We analyze the Abelian-Higgs Model in the gravity background which is dual to thermal
non-relativistic ( NR ) conformal field.

In the present work, we find that there is also a critical temperature like the relativistic case, below which a
charged condensation field appears by a second order phase transition and the (DC) conductivity becomes infinite. In
particular, we find that as the non-relativistic parameter increases, the condensation happens more observably. We also
calculated the frequency dependent conductivity and find that as the non-relativistic parameter increases, the transition happens more observably and frequency
positions of peaks move away from the $\omega=0$ axis.

\section{Holographic Abelian Higgs model in non-relativistic regime}
\subsection{Gravity for NR conformal field}
AdS/CFT correspondence has been extended to describe
non-relativistic condensed matter system recently
~\cite{Herzog:2008wg,Son:2008ye,Adams:2008wt}. In the present work,
we start with the  the gravity background with a black hole coming
from Null Melvin Twist of the planar Schwarzschild anti-de Sitter
black hole ~\cite{Herzog:2008wg}
\begin{eqnarray} \dd s_{\textrm{Einstein}}^2&=& r^2\,k^{-\frac{2}{3}} \,\biggr[ -\beta^2 \,r^2 \,f \,(\dd t+\dd y)^2 - f \,\dd t^2 + \dd y^2 + k
\,\dd {\bf x}^2 \biggr] +  k^{\frac{1}{3}}\, \frac{\dd r^2}{r^2\, f} \ . \nonumber
\end{eqnarray}
In light cone coordinates, the above metric turns to be
\begin{eqnarray}\label{metric}
 \dd s^2&=& r^2\,
k^{-\frac{2}{3}}\biggr[\left(\frac{1-f}{4\beta^2} -
    r^2\,f\right) \, \dd u^2 + \frac{\beta^2 r_+^4}{r^4} \, \dd v^2 - \left(1+f\right)\,\dd u\,\dd v \biggr] \nonumber \\
&& \quad+\;\;  k^{\frac{1}{3}}\, \left(r^2 \dd {\bf x}^2 +  \frac{\dd r^2}{r^2\, f} \right),
\end{eqnarray} where \be f = 1- \frac{r_+^4}{r^4} \ , \qquad k = 1 +\beta^2 \,r^2 \,(1-f)  = 1 + \frac{\beta^2
\, r_+^4}{r^2}. \ee  The light cone coordinates are
\begin{equation}
u = \beta \, (t+y) \ , \qquad v = \frac{1}{2\,\beta} \, (t-y). \label{rescaleuv}
\end{equation}
$\beta$ is a non-relativistic parameter and we choose $\beta=1$ in the present work if no specification. $r_+$
determines the Hawking temperature of the black hole ~\cite{Herzog:2008wg} \be T = \frac{r_+}{\pi \,\beta}\,. \ee This
black hole is 4+1 dimensional, and so will be dual to a 2+1 dimensional non relativistic theory. The metric
(\ref{metric}) has asymtotic Schr$\ddot{\textrm{o}}$dinger symmetry, which can be easily found if we set $r_+=0$. By
the view of gauge/gravity, this black hole can be expected to describe non-relativistic strongly correlated quantum
criticality. We shall use some probe fields to uncover the transition in the following subsection.

\subsection{The model}
We start with the background of a black hole with asymtotic Schr$\ddot{\textrm{o}}$dinger symmetry (\ref{metric})
\begin{eqnarray}
 \dd s^2=r^2\,
k^{-\frac{2}{3}}\biggr[(\frac{1-f}{4\beta^2} -
    r^2f) \dd u^2 + \frac{\beta^2 r_+^4}{r^4} \dd v^2 - (1+f)\dd u\dd v \biggr] +  k^{\frac{1}{3}} (r^2 \dd {\bf x}^2 +  \frac{\dd r^2}{r^2 f}
    )\,.
\end{eqnarray}
In this background, we now consider a Maxwell field and a charged complex scalar field. The action turns to
be\footnote{ Setting $g=1$ is a choice of units of charge in the dual 2+1 theory ~\cite{Hartnoll:2008vx}. }
\be\label{eq:action} S =\int \dd^5x \sqrt{-g}\biggr[- \frac{1}{4} F^{ab} F_{ab} - V(|\Psi|) - |\pa \Psi - i A \Psi |^2
\,\biggr].\ee
For simplicity and concreteness, we choose the quadratic potential
and ignore the higher terms\be\label{eq:mass} V(|\Psi|) =
2\,\widetilde{a} |\Psi|^2\ , \ee where $\widetilde{a}$ is a negative
constant parametrizing the symmetry breaking.
 Due to plane symmetric ansatz, $\Psi = \Psi(r)=\psi$,
equation of motion of the scalar field is
\bea\label{psieqa}\begin{split} \biggr\{-8\widetilde{a}r^7f \beta^2
k^{1/3}+r\biggr[4r^4(2-r^2)\beta^2+r_+^4(1+4(r^2-1)\beta^2+4\beta^4)
\biggr] \phi^2 \biggr\}\psi\qquad \\
+4r^4f\beta^2\left[r^4(4+f)\psi^\prime+r^5f\psi^{\prime\prime}
\right]=0,
\end{split}\eea where the scalar
potential $\phi$  is the  electric potential in the axial gauge
$A_i=0$ so that $A_u \doteq A_v =\phi(r)$ . \footnote{In the light cone
coordinates $u$ and $v$, the original electronic field $A_t$ has been
transformed to $A_u$ and $A_v$. Our argument is $A_v$ can be regarded as a assistant field contributing the electric potential and it has no independent Maxwell equation of motion.} We shall rewrite the Lagrangian with $\psi$ and $\phi$, thus the equation for the scalar
potential $\phi$ is \bea\label{A0eqa}
\begin{split}
&6k^{1/3} \phi \psi ^2 \biggr\{r_+^8 \biggr[4
   \left(r^2-1\right) \beta ^4+4 \beta ^6+\beta
   ^2\biggr]-4 r^6 \left(r^2-2\right) \beta ^2 \\
   &+r^2 r_+^4
   \biggr[4 \left(r^2-1\right) \beta ^2+\left(-4 r^4+8
   r^2+4\right) \beta ^4+1\biggr]\biggr\}\\
   &+f r \biggr\{3 r \biggr[-r_+^8 \left(4
   \left(r^2-1\right) \beta ^4+4 \beta ^6+\beta
   ^2\right)+4 r^6 \left(r^2-2\right) \beta ^2\\
   &+r^2
   r_+^4 \left(-4 \left(r^2-1\right) \beta ^2+4
   \left(r^4-2 r^2-1\right) \beta ^4-1\right)\biggr] \phi
   ''\\
   &+\biggr[r_+^8 \left(-4 \left(5 r^2+1\right) \beta
   ^4+4 \beta ^6+\beta ^2\right) +12 r^6 \left(5
   r^2-6\right) \beta ^2 \\
   &+r^2 r_+^4 \left(-12
   \left(r^2+1\right) \beta ^2+4 \left(17 r^4-22
   r^2+3\right) \beta ^4+3\right)\biggr] \phi '\biggr\}= 0,
\end{split}\eea where $\psi^2$ is the, in our case,
$r$ dependent mass. The charged condensate has triggered a Higgs mechanism in the gravity background.

\section{Numerical results}
\subsection{Condensation}
To compute the expectation value of operators in dual field theory,
we need  the asymptotic behaviors of equations (\ref{psieqa}) and
(\ref{A0eqa}). In the limit  $r\rightarrow\infty$, we find the
equations of motion for $\psi$ and $\phi$ become\bea
\begin{split}
r^2\psi^{\prime\prime}+5r\psi^{\prime}-(2\widetilde{a}+\phi^2)\psi  &=&0,\\
r^2\phi^{\prime\prime}+5r\phi^{\prime}-2\psi^2\phi  &=& 0,
\end{split}\eea
where we can define a new quantity
$a=\widetilde{a}+\frac{1}{2}\mu^2$. The asymptotic solution of equation (\ref{psieqa}) is \bea
\begin{split}\psi \rightarrow r^{-\lambda_+} C_1 + r^{-\lambda_-}
C_2,
\end{split}\eea
with   $\lambda_\pm=2\pm \sqrt{4+2a}$ and the asymptotic behavior of (\ref{A0eqa}) is \bea\label{A0eqa1}
\phi\rightarrow \mu-{\rho\ov r^4}\,,
 \eea  where $\mu$ and $\rho$ are chemical potential and charge density.
  By AdS/CFT correspondence, we can interpret the coefficients $C_\pm$ as the expectation values of the
 operator $O_\pm$ whose conformal dimension is $\lambda_\pm$ respectively.
  We will fix one of them, and compute the other.
\begin{figure}[h]
\begin{center}
\epsfig{file=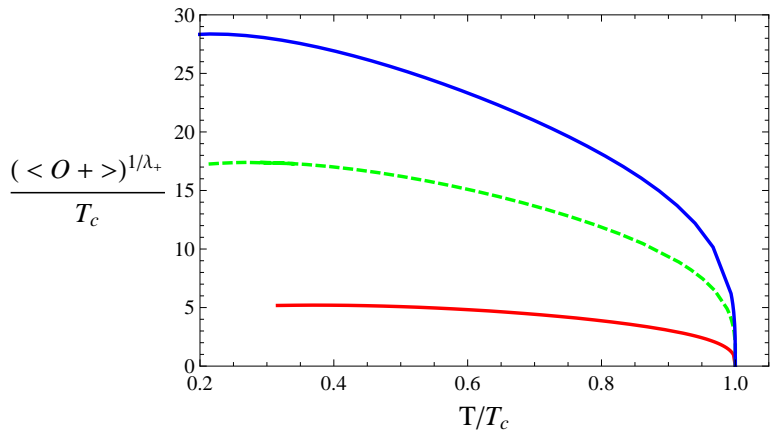,width=4in,angle=0,trim=0 0 0 0}
\end{center}
\caption{We plot expectation values of the operators as a function of temperature. As $\lambda_+=2+ \sqrt{4+2a}$
increases, the condensation increases. The red real line is for $a=-1.5$, the green line is for $a=-1.2$ and the blue
real line is for $a=-1$.
  \label{fig:psi1}}
\end{figure}

\begin{figure}[h]
\begin{center}
\epsfig{file=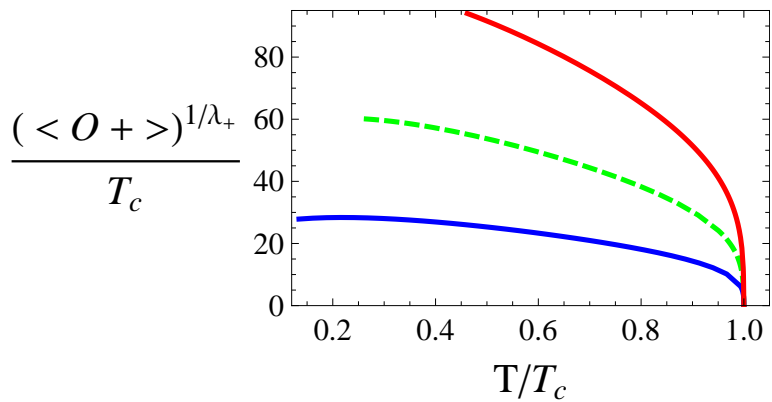,width=4in,angle=0,trim=0 0 0 0}%
\end{center}
\caption{We focus on condensation in $a=-1$ case and  show them with
different non-relativistic parameter $\beta$. The blue line is for
$\beta=1$, the green dashed line is for $\beta=1.5$ and the red line
is for $\beta=2$.
  \label{fig:psi2}}
\end{figure}

In figure \ref{fig:psi1}, we show the behavior of the expect values of condensation field as the temperature goes down.
We find that below a critical temperature $T_c$, the condensation field appears and obtains finite value. As argued in
~\cite{Horowitz:2008bn}, here increasing $\lambda_+$ corresponds to increasing the mass of the bulk scalar. This figure
shows that as the condensation field becomes heavier, the transition happens more observably. In figure \ref{fig:psi2},
it shows that, as the non-relativistic parameter becomes larger, the condensation happens more observably.
We can find difference between non-relativistic case and relativistic case from this figure. As shown in
~\cite{Herzog:2008wg}, $\beta$ is proportional to Galilean mass in non-relativistic system, then we can see that
non-relativistic particles with a heavier Galilean mass can condense more observably. We argue that a very small
$\beta$ in non-relativistic can be closed to relativistic case.

We find the behavior of order parameter near the critical temperature with different $a$ and $\beta$ can be described
by an universal exponent
\begin{equation}
<O_+>\; \varpropto \; T_c(1-T/T_c)^\xi
\end{equation}
where $\xi\approx1/3$.

\subsection{Conductivity}

We shall compute the conductivity in the dual conformal field theory as a function of frequency. As first, we want to
solve fluctuations of the vector potential $A_x$ in the bulk.  The Maxwell equation at zero spatial momentum and with a
time dependence of the form $e^{- i \w u}$ gives \bea\begin{split} 3 r \left(r^2+r_+^4\right) \biggr[-r_+^4 \omega ^2+2
r^2 \left(r^4-r_+^4\right) \left(1+\frac{r_+^4}{r^2}\right)^{1/3} \psi^2\biggr]A_x&\\
-\left(r^4-r_+^4\right) \biggr[\left(9 r^6+3 r^2 r_+^4+11 r^4 r_+^4+r_+^8\right) A_x '&\\
+3 r \left(r^6-r^2 r_+^4+r^4 r_+^4-r_+^8\right) A_x ''\biggr]& = 0\,.\end{split}\eea To compute the retarded (causal) green function, we
solve this equation with ingoing wave boundary conditions at the horizon \cite{Son:2002sd}: \be\label{Axeom} A_x
\propto (r^4-r_+^4)^{-{i\omega\over 4 r_+}}.\ee The asymptotic behavior of the Maxwell field at large radius turns to
be \be\label{aseomAx} A_x = A_x^{(0)} + \frac{A_x^{(1)}}{r^2} + \cdots \ee In a rough way, the AdS/CFT dictionary tells
us that the dual source and expectation value for the current are given by \be A_x = A_x^{(0)} \,, \qquad \langle J_x
\rangle = A_x^{(1)} \,. \ee Now from Ohm's law we can obtain the conductivity \be\label{sigma} \sigma(\w) =
\frac{\langle J_x \rangle}{E_x} = - \frac{ \langle J_x \rangle}{\dot A_x} = -\frac{ i \langle J_x \rangle}{\w A_x} = -
\frac{i A_x^{(1)}}{\w A_x^{(0)}} \,. \ee More concretely, we can calculate the conductivity from the correlation
function. The term in the action which contains two derivatives with respect to $r$ is \bea\begin{split}
S&=-{1\ov 4}\int \dd r\dd^4 x\sqrt{-g}g^{rr}g^{xx}\partial_rA_x\partial_rA_x+\cdots\\
&=\int \dd r\dd^4 x F(r)\partial_rA_x\partial_rA_x+\cdots
\end{split}
\eea The retarded Green function in Minkovski space is\footnote{More details about analysis of correlation function can
be found in the Appendix A.} \be
 G_R=-F(r)A_xA_x'\biggr|_{r\rightarrow +\infty}\,,
\ee where \be F(r)={fr^3\ov 2k^{1/3}}\,. \ee The conductivity is given by
\be
 \sigma(\omega)={1\ov i\omega}G_R={-F(r)A_xA_x'\ov i\omega}\biggr|_{r\rightarrow +\infty}.
\ee From (\ref{Axeom}), we can see that \be \sigma(\omega)={A_{(1)}\ov i\omega A_{(0)}} \,,\ee which is same as the result in
(\ref{sigma}).
\begin{figure}[h]
\begin{center}
\epsfig{file=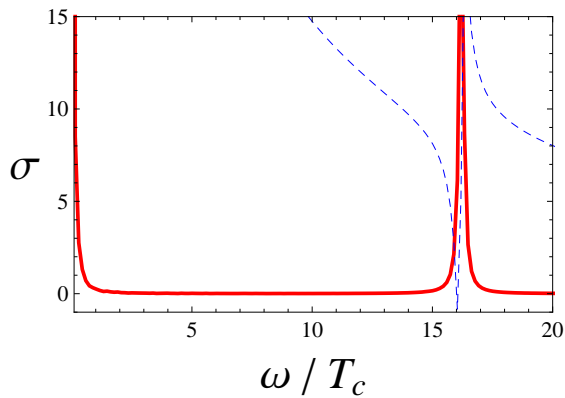,width=2.9in,angle=0,trim=0 0 0 0}%
\hspace{0.3cm}
\end{center}
\caption{A.C  conductivity for NR superconductors. Each plot is at
the temperature below the critical temperature $T/T_c\approx0.1$.
The red line is the real part and the blue line is the imaginary
part. In this figure, $a=-0.5$.
  \label{fig:gap3}}
\end{figure}

\begin{figure}[h]
\begin{center}
\epsfig{file=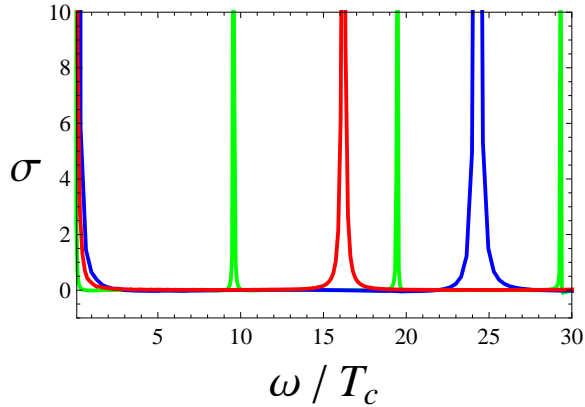,width=3in,angle=0,trim=0 0 0 0}%
\end{center}
\caption{Conductivity for NR superconductors. In this figure, for
fixed $a=-0.5$, we can see real parts of conductivities with
different non-relativistic parameter $\beta$. The blue, red and
green line corresponds to $\beta=1.5$, $\beta=1$, $\beta=0.5$
respectively.
  \label{fig:4}}
\end{figure}

In figure \ref{fig:gap3}, the real part of $\sigma(\omega)$ is infinity at $\omega=0$, as shown in the figure
apparently. We can also see this pole at $\omega=0$ from the imaginary part.
We have multiple peaks for the conductivity. These peaks show that there are also more frequency modes which have
relative large conductivities. Similar results for relativistic superconductor have been found in
~\cite{Horowitz:2008bn}~\cite{Ammon:2009fe}, but so far we have no direct physical interpreting for these peaks. From
numerical process, these peaks comes from terms of high powers of $r$ in EOM of $A_x(r)$, which is the same as the case
in ~\cite{Ammon:2009fe}, and there similar peaks have also been interpreted as quasi-particle excitations. In our case,
we can consider that $A_x$ is not just a perturbation but a heavy field, then its corresponding boundary vector
excitations can appear. We can calculate the spectral function by $-2 \textrm{Im}G$ . In our model, these peaks may be
such vector quasi-particles.

Actually, when we increase $a$, we find the peaks move to $\omega=0$
axis. We also computed the difference among conductivities with
different non-relativistic parameter $\beta$ in the Figure
\ref{fig:4}. We find that, as $\beta$ decreases, the frequency
positions of peaks move to the $\omega=0$ axis. It seems reasonable
when we consider that $\beta$ reflects the mass of non-relativistic
particles, for a bigger $\beta$, the transition happens more
observably and peaks are farther from the $\omega=0$ axis. \footnote{More
details about $\beta$ dependence can be found in the Appendix B.}

\begin{figure}[h]
\begin{center}
\epsfig{file=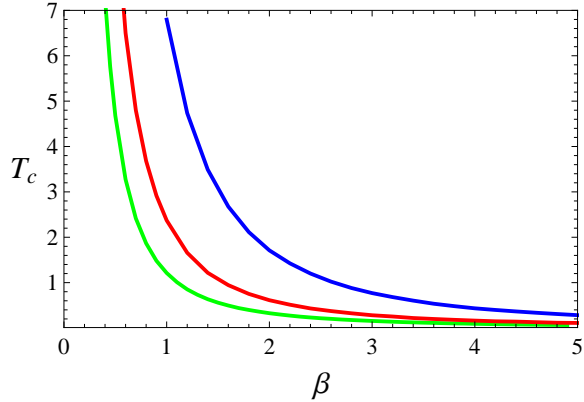,width=3in,angle=0,trim=0 0 0 0}%
\end{center}
\caption{Critical temperature is plotted as a function of $\beta$. In this figure, the critical temperature decreases
as non-relativistic parameter $\beta$ increases. The blue, red and green lines correspond to $a=-0.5$, $a=-1$ and
$a=-1.5$ respectively.
  \label{fig:5}}
\end{figure}

In figure \ref{fig:5}, we observe that the critical temperature will decrease as non-relativistic parameter $\beta$
increases. It implies that as the non-relativistic particle is heavier, the transition temperature turns smaller. Note
that $\beta$ contributes the mass of fundamental particles, not the mass of the quasi-particles in condensed matter
system.

\section{Conclusion}
In the present work, we built a  holographic  model for the strongly interacting non-relativistic system  showing
superconductivity. We used a black hole background which satisfies an asymptotic Schrodinger symmetry to describe the
non-relativistic strongly correlated condensed matter system. We introduced a two components $A_u=A_v=\phi$
 field to describe the electric field and a complex scalar to describe the condensed field.  There is also a critical
temperature like the relativistic case, below which a charged condensate field appears by a second order phase
transition and the (DC) conductivity becomes infinite. In particular, we find that as the non-relativistic parameter
increases, the condensation happens more observably. We also calculated the frequency dependent conductivity and find
that as the non-relativistic parameter decreases, frequency positions of peaks move to the $\omega=0$ axis.

After we almost finished the present work, we note that the work~\cite{Liu:2009dm} extends the holographic model to the
phase which covers fermions, like the phase before superconductor transition. Two such fermions may bound a
quasi-particle to interpret our peaks. This model can give interesting result like fermi-surface for the non-fermi
liquid. And the work~\cite{Hartnoll:2009sz} gives more concrete program to describe condensed matter system
holographically. All these methods can be easily used for relativistic critical phenomena. However, constructing
holographic non-relativistic model for critical phenomena is still an unsolvable problem since we can not easily get a
good asymptotic behavior of bulk field in the non-relativistic gravity background (after light cone transformation of
the original coordinates). We introduce an extra field component in $v$ direction for electric field to solve this
problem and give an example for the non-relativistic superconductor. More details about non-relativistic holographic
construction for the strongly correlated system need to investigate.

Before the present letter, many works about holographic quantum critical system appeared. We hope similar methods can
be used for the better construction of holographic model of the non-relativistic quantum critical phenomena in the
future.

\section*{Acknowledgments}
We thank  Miao Li, Shuo Yang, Yumi Ko and Xu-guang Huang for useful
discussions. This work is supported by KOSEF Grant
R01-2007-000-10214-0 and also by the SRC Program of the KOSEF with
grant number R11 - 2005 -021.

\section*{Appendix}
\subsection*{A. calculation of conductivity}
The Maxwell field $A=A_\mu \dd x^\mu$, where\be A_\mu=(0,0,A_x,0,0)\,, \ee \be x^\mu=(t, r, x, y, v)\,.\ee The metric
$G$ is given by\footnote{In the appendix, $t$ is actually $u$ in main body of this paper.}
\begin{center}
 $G_{5\times 5}(g_{\mu\nu}) = \left(\begin{array}{ccccc}
r^2k^{-2/3}({1-f\ov 4\beta^2}-r^2f) & 0 & 0 & 0 & -{r^2\ov 2}k^{-2/3}(1+f) \\
0 & {k^{1/3}\ov r^2f} & 0 & 0 & 0 \\
0 & 0 & r^2k^{1/3} & 0 & 0 \\
0 & 0 & 0 & r^2k^{1/3} & 0 \\
-{r^2\ov 2}k^{-2/3}(1+f) & 0 & 0 & 0 & r^2k^{-2/3}\beta^2{r_0^4\ov r^4}
\end{array}\right)\,,$
\end{center}
where \be f=1-{r_0^4\ov r^4},\quad k=1+\beta^2{r_0^4\ov r^2}\,.\ee Then the inverse metric $G^{-1}$ is
\begin{center}

$G_{5\times 5}^{-1}(g^{\mu\nu}) = \left(\begin{array}{ccccc}
{(1-f)k^{2/3}\ov ((f-1)r^2-1)r^2f} & 0 & 0 & 0 & {k^{2/3}(1+f)\ov 2((f-1)r^2-1)r^2f} \\
0 & {r^2f\ov k^{1/3}} & 0 & 0 & 0 \\
0 & 0 & {1\ov r^2k^{1/3}} & 0 & 0 \\
0 & 0 & 0 & {1\ov r^2k^{1/3}} & 0 \\
{k^{2/3}(1+f)\ov 2((f-1)r^2-1)r^2f} & 0 & 0 & 0 & {(1-f-4fr^2)k^{2/3}\ov 4((f-1)r^2-1)r^2f}
\end{array}\right)\,.$
\end{center}
The action with term including $A_x$ is given by \bea\begin{split} S=\int \dd^4 x\dd r\sqrt{-g}&\left(-{1\ov
4}F^2-g^{\mu\nu}A_\mu A_\nu |\psi|^2\right)\\
=\int \dd^4 x\dd r\sqrt{-g}&\biggr[-{1\ov 2}(g^{rr}g^{xx}\p_rA_x\p_rA_x + g^{tt}g^{xx}\p_tA_x\p_tA_x \\
&+g^{yy}g^{xx}\p_yA_x\p_yA_x + 2g^{tv}g^{xx}\p_tA_x\p_vA_x)\\
&-g^{xx}A_x A_x |\psi|^2\biggr]\,.
\end{split}\eea
We expend the $A_x$\footnote{$\phi$ is $A_x(r)$ in Section 2, which is different from the $\phi$ in Section 2.}
\bea\label{Axextend}\begin{split}
 A_x&=\int \dd \omega \dd p_1\dd p_2 \dd p_3 e^{-i\omega t+ip_1y+ip_2x+ip_3v}\phi_{\omega,p_1,p_2,p_3}(r)\varphi(p_1,p_2,\omega)\delta_{p_3,k_3}\\
&=\int \dd^4k  e^{-i\vec k \vec x+ip_3v}\phi_{\vec k,p_3}(r)\varphi(\vec k)\delta_{p_3,k_3}\,,
 \end{split}\eea where $k_3$ is considered as a constant. We put the (\ref{Axextend}) into the action and obtain the five
 parts of the action\footnote{$\delta_{p_3,-p_3'}$ here just gives a constraint for the action in momentum space.}
 \bea\begin{split} S_1&=-\delta_{k_3,-k_3'}\int \dd^3 k\int_{r_H}^{r_B}\dd r {1\ov
2}\sqrt{-g}g^{rr}g^{xx}\phi_{\vec k}'(r)\phi_{-\vec k}'(r)\varphi(\vec k)\varphi(-\vec k )\\
S_2&=-\delta_{k_3,-k_3'}\int \dd^3 k\int_{r_H}^{r_B}\dd r {\omega^2\ov 2}\sqrt{-g}g^{tt}g^{xx}\phi_{\vec
k}(r)\phi_{-\vec
k}(r)\varphi(\vec k)\varphi(-\vec k)\\
S_3&=-\delta_{k_3,-k_3'}\int \dd^3 k\int_{r_H}^{r_B}\dd r {p_1^2\ov 2}\sqrt{-g}g^{yy}g^{xx}\phi_{\vec k}(r)\phi_{-\vec
k}(r)\varphi(\vec k)\varphi(-\vec k)\\
S_4&=-\delta_{k_3,-k_3'}\int \dd^3 k\int_{r_H}^{r_B}\dd r {\omega p_3}\sqrt{-g}g^{tv}g^{xx}\phi_{\vec
k}(r)\phi_{-\vec k}(r)\varphi(\vec k)\varphi(-\vec k)\\
S_5&=-\delta_{k_3,-k_3'}\int \dd^3 k\int_{r_H}^{r_B}\dd r \sqrt{-g}g^{xx}|\psi|^2\phi_{\vec k}(r)\phi_{-\vec
k}(r)\varphi(\vec k)\varphi(-\vec k)\,.
\end{split}\eea
We assume \be \phi_{\vec k,p_3}(r_B)=1, \ee then the retarded Green function ~\cite{Son:2002sd} is  \bea\begin{split}
G_{xx}^R(\vec k,p_3)=\delta_{k_3,-k_3'}\biggr\{\int\dd r &\biggr[-{1\ov
2}\sqrt{-g}g^{rr}g^{xx}\phi_{\vec k}'(r)\phi_{-\vec k}'(r)\\
&-{\omega^2\ov 2}\sqrt{-g}g^{tt}g^{xx}\phi_{\vec k}(r)\phi_{-\vec
k}(r)\\
&-{p_1^2\ov 2}\sqrt{-g}g^{yy}g^{xx}\phi_{\vec k}(r)\phi_{-\vec
k}(r)\\
&-{\omega p_3}\sqrt{-g}g^{tv}g^{xx}\phi_{\vec
k}(r)\phi_{-\vec k}(r)\\
&-\sqrt{-g}g^{xx}|\psi|^2\phi_{\vec k}(r)\phi_{-\vec k}(r)\biggr]\biggr\}\biggr|^{r_B}\,.
\end{split}\eea
We set $r_B\rightarrow +\infty$, then we simplify the above result \bea\label{Gf}\begin{split} G_{xx}^R(\vec
k,p_3)=\delta_{k_3,-k_3'}\biggr\{\int\dd r &\biggr[-{1\ov
2}r^3\phi_{\vec k}'(r)\phi_{-\vec k}'(r)\\
&+{\omega^2\ov 2}{r_+^4\ov r^5}\phi_{\vec k}(r)\phi_{-\vec
k}(r)\\
&-{p_1^2\ov 2}{1\ov r}\phi_{\vec k}(r)\phi_{-\vec
k}(r)\\
&+{\omega p_3}{1\ov r}\phi_{\vec
k}(r)\phi_{-\vec k}(r)\\
&-r|\psi|^2\phi_{\vec k}(r)\phi_{-\vec k}(r)\biggr]\biggr\}\biggr|^{r_B}\,,
\end{split}\eea
when $r\rightarrow +\infty$, from asymtotical solution (\ref{aseomAx}) of $A_x$, the leading terms are the following
\bea\label{greenfun}\begin{split} G_{xx}^R(\vec k,p_3)=\delta_{k_3,-k_3'}\biggr[\int\dd r &-{1\ov 2}r^3\phi_{\vec
k}'(r)\phi_{-\vec k}'(r)\biggr]\biggr|^{r_B}\,,
\end{split}\eea
when the condensation field is zero, we can quickly give the Green function \bea\begin{split} G_{xx}^R(\vec
k,p_3)=\lim_{r\rightarrow +\infty}-{1\ov 2}r^3\phi_{\vec k}(r)\phi_{-\vec k}'(r)\,. \end{split}\eea From Kubo formular,
we see the conductivity \be \sigma(\omega)={1\ov i\omega}G^R = \lim_{r\rightarrow +\infty}-{1\ov 2i\omega}r^3\phi_{\vec
k}(r)\phi_{-\vec k}'(r) \ee When temperature is below a critical value and $|\psi|$ has finite value, then the last
term in (\ref{Gf}) tells us that the conductivity becomes large.
\subsection*{B. $\beta$ dependence of condensation and conductivity}
To compare the non-relativistic case with relativistic case, we focus on the $\beta$ dependence of the condensation
field and the conductivity. From Figure \ref{fig:psi2}, we can see that as $\beta$ increases, the condensation happens
and condensed field appears more observably. To compute the conductivity, the equation of motion including $\beta$
turns to be \bea\begin{split} 3 r \left(r^2+r_+^4\beta^2\right) \biggr[-r_+^4 \beta^2\omega ^2+2
r^2 \left(r^4-r_+^4\right) \left(1+\frac{r_+^4}{r^2}\right)^{1/3} \psi^2\biggr]A_x&\\
-\left(r^4-r_+^4\right) \biggr[\left(9 r^6+3 r^2 r_+^4+11 r^4 r_+^4\beta^2+r_+^8\beta^2\right) A_x '&\\
+3 r \left(r^6-r^2 r_+^4+r^4 r_+^4\beta^2-r_+^8\beta^2\right) A_x ''\biggr]& = 0\,.\end{split}\eea Then the near
horizon solution is given by \be A_x \propto (r^4-r_+^4)^{-{i\omega\beta\over 4 r_+}}.\ee And the conductivities with
different $\beta$ are given by Figure \ref{fig:4}. We find that, as $\beta$ increases, the frequency positions of peaks
move to the $\omega=0$ axis. We argue it is reasonable, since when we consider that the $\beta$ reflects the mass of
the condensation field in non-relativistic case, for a larger $\beta$, the transition happens more observably and peaks
move to the $\omega=0$ axis.

\end{document}